\documentclass[preprint]{elsarticle}
\usepackage{amsmath}
\usepackage{amsfonts}
\usepackage{amscd,amsthm}
\usepackage{graphicx}
\usepackage{amssymb}

\theoremstyle{plain}

\newtheorem{theorem}{Theorem}[section]

\newtheorem{remark}[theorem]{Remark}

\newcommand{\mc}{\mathcal}

\begin{document}

\date{\today}

\title{Improving beam stability in particle accelerator models by using Hamiltonian control}

\author[jb]{J.~Boreux\corref{cor1}}
\ead{jehan.boreux@fundp.ac.be}

\author[jb]{T.~Carletti}
\ead{timoteo.carletti@fundp.ac.be}

\author[chs]{Ch.~Skokos}
\ead{hskokos@pks.mpg.de}

\author[mv]{M.~Vittot}
\ead{vittot@cpt.univ-mrs.fr}

\cortext[cor1]{Corresponding author}

\address[jb]{Namur Center for Complex Systems, naXys,
University of Namur\\
Rempart de la Vierge 8, Namur, 5000, Belgium}

\address[chs]{Max Planck Institute for the Physics of
Complex Systems\\
N\"othnitzer Str. 38 D-01187, Dresden, Germany}

\address[mv]{Centre de Physique Th\'eorique CNRS Luminy\\
case 907 - 13288 Marseille cedex 9, France}

\begin{abstract}
We derive a Hamiltonian control theory which can be applied to a 4D symplectic
map that models a ring particle accelerator composed of elements with sextupole
nonlinearity. The controlled system is designed to exhibit a more regular
orbital behavior than the uncontrolled one. Using the Smaller Alignement Index
(SALI) chaos indicator, we are able to show that the controlled system has a
dynamical aperture up to $1.7$ times larger than the original model.
\end{abstract}

\begin{keyword}
hamiltonian system \sep chaos \sep control \sep accelerator mappings \sep dynamical aperture
\end{keyword}

\maketitle
\section{Introduction}
\label{sec:intro}

Particle accelerators are technological devices which allow studies at both
\lq\lq infinitely small scale\rq\rq, {e.g.~particles} responsible for
elementary forces, and \lq\lq extremely large scale\rq\rq, e.g.~the
origin of cosmos. In a simplified approach, such devices are composed of basic elements sequence: focusing and defocusing magnets,
accelerating electromagnetic fields and trajectory bending elements as they are used in the
case of ring accelerators. The resulting dynamics is nonlinear, and can be
described, in the absence of strong damping, by a conservative system. This system 
 can be modelled by a symplectic map built from the composition 
of several elementary maps corresponding to each basic magnetic element.

One of the main problems found in the dynamics of ring {accelerators} is to study
the stability around the nominal orbit, {i.e.~the} circular orbit passing through
the centre of the ring. Each component of the ring can be seen as a
nonlinear map, that deforms the trajectory at large amplitude. Moreover, such maps possess
stochastic layers whose effect is the reduction of the stability domains
around the nominal circular orbit (the so-called {\em dynamical aperture} -
DA) \cite{GST1997}. Such behaviors imply that (chaotic) nearby orbits can drift away
after a few ring {turns}, eventually colliding with accelerator's boundaries,
and consequently reduce the beam lifetime and the performance of the accelerator.

The aim of the present paper is to derive a reliable improvement of the
stability of the beam by increasing the DA in a simplified accelerator model, consisting
of only one type of element having a sextupole
nonlinearity \cite{BT1991,BK1994,VBK1996,VIB1997}.

We work in the framework of the {\em Hamiltonian Control Theory} presented
in~\cite{V2004,CVECP2005}, where two methods to control symplectic maps
have been described, namely using Lie transformations and generating
functions. In the present paper we use the former method, that allows 
direct determination of the new controlled map; avoiding the possible problems
related to coordinate inversion.

The aim of control theory is to improve selected features of a given
system, by slightly modifying its Hamiltonian with the addition of a {\em
small} control term, so that the new system and the unperturbed one are
conjugated namely, they have the same dynamics. This technique is
particularly suitable whenever one can directly act on the system and modify
it, e.g.~in the case of a particle accelerator where the addition
of a control term in the Hamiltonian function can be seen as the introduction
of a suitable magnet in the accelerator lattice.

In our study, we use the Smaller Alignement Index
(SALI)~\cite{BS2006,S01,SABV03,SABV04} method, which is an efficient indicator
for characterising orbits as chaotic or regular in Hamiltonian flows and
symplectic maps. The SALI is computed using the time
evolution of two deviation vectors along the studied orbit.

The paper is organized as follows: we introduce the model in Section~\ref{sec:themodel} and present a general result for the
control of symplectic maps in Section~\ref{sec:ctrlth}. We apply the theory to the symplectic map model of a standard ring accelerator {in} Section~\ref{sec:fodoctrl}. We briefly recall the
SALI chaos indicator in Section~\ref{sec:salimeth}, while
Section~\ref{sec:results} presents our numerical results on the behavior of the constructed model. {Finally, in Section~\ref{sec:ccl} we summarize our
conclusions.}  Further technical details can be found in the Appendix.

\section{The model}
\label{sec:themodel}

Let us consider a system consisting of a charged particle and a simplified accelerator
ring with linear frequencies (tunes) $q_x$, $q_y$, with a localized thin
sextupole magnet {(for more details the interested reader is referred
to~\cite{BT1991})}. The magnetic field {of this element} 
modifies the orbit once the particle passes through it.
The basic model is :
\begin{equation}
  \label{eq:4DFODO}
  \left(
    \begin{smallmatrix}
      x_1^{\prime}\\x_2^{\prime}\\x_3^{\prime}\\x_4^{\prime}
    \end{smallmatrix}
\right)=\left(
  \begin{smallmatrix}
    \cos \omega_1 & -\sin\omega_1 & 0 & 0 \\
    \sin \omega_1 & \cos\omega_1 & 0 & 0 \\
    0 & 0 & \cos \omega_2 & -\sin\omega_2 \\
    0 & 0 & \sin \omega_2 & \cos\omega_2 \\
  \end{smallmatrix}
\right)  \left(
    \begin{smallmatrix}
      x_1\\x_2+x_1^2-x_3^2\\x_3\\x_4-2x_1x_3
    \end{smallmatrix}
\right)=T\left(
    \begin{smallmatrix}
      x_1\\x_2\\x_3\\x_4
    \end{smallmatrix}
\right)\, 
\end{equation}
where $x_1$ {($x_3$)} denote the {deflection} from the ideal circular orbit
in the horizontal {(vertical)} direction before the particle enters the
element and $x_2$ {($x_4$) are the associated momentum.} Primed variables denote
positions and momenta after the particle left the element.  The
parameters $\omega_1$ and $\omega_2$ are related to the accelerator's
tunes~\footnote{Such parameters have been fixed throughout this work to the
values $q_x=0.61803$ and $q_y=0.4152$, corresponding to a non-resonant
condition (see~\cite{VIB1997}).} $q_x$ and $q_y$ by the relations $\omega_1 = 2\pi q_x$ and
$\omega_2 = 2\pi q_y$.  The first matrix in (\ref{eq:4DFODO}) describes the
linear motion of a particle, which corresponds to a simple rotation in the
phase space. The nonlinearity induced by the thin sextupole magnet is
modeled by the 2$^{nd}$ order polynomial expression in
(\ref{eq:4DFODO}). The particle dynamics at the $n$-th turn,
can be described by the sequence
$(x_1^{(n)},x_2^{(n)},x_3^{(n)},x_4^{(n)})_{n\geq 0}$, where the $(n+1)$-th
positions and momenta are defined as a function of the $n$-th ones
by~\eqref{eq:4DFODO}.

{The map (\ref{eq:4DFODO})} decomposes in an integrable part and a quadratic perturbation, respectively the system is associated to
 the following {Hamiltonian} 
(see~\ref{sec:appendix})
\begin{equation}
  \label{eq:hamrot}
  H(x_1,x_2,x_3,x_4)=-\omega_1\frac{x_1^2+x_2^2}{2}-\omega_2\frac{x_3^2+x_4^2}{2}\quad
  \text{and} \quad V(x_1,x_2,x_3,x_4)=-\frac{x_1^3}{3}+x_1x_3^2\, ,
\end{equation}
more precisely (\ref{eq:4DFODO}) can be written in terms of Poisson brackets~\footnote{In the
literature one can sometimes find the alternative equivalent notation $\{ H\}
= L_H$.} as
\begin{equation}
\vec{x}^{\prime}=T(\vec{x})=e^{\{H\}}e^{\{V\}}\vec{x}\, .
\label{eq:pertflow}
\end{equation}
Here $\vec{x}=(x_1,x_2,x_3,x_4)^{\rm T}$, {with $^{\rm T}$ denoting the
transpose of a matrix,} and by definition, for any function $f$ defined in the
phase space, $\{H\}f =\{H,f\}=(\nabla H)^{\mathrm T} J\nabla f${, with
$J=\left( \begin{array}{cc} \mathbf{0} & \mathbf{1} \\ -\mathbf{1} &
\mathbf{0}
\end{array}
 \right) $}, being  the symplectic constant matrix,
\begin{equation}
e^{\{H\}}f=\sum_{n\geq 0}\frac{\{H\}^n}{n!}f\quad \text{and}\quad
\{H\}^nf=\{H\}^{n-1}(\{ H\}f)\, .
\label{eq:exp}
\end{equation}

Maps of the form~\eqref{eq:4DFODO} have already been studied {in
~\cite{BS2006} where it has been shown} that chaotic orbits reduce the DA to a hypersphere
of radius $\sim 0.39$ in the 4-dimensional phase space (see Figs.~5 and 6
of~\cite{BS2006}). The goal of {the present paper} is to show that the
stability region of the nominal circular orbit {can be increased} once the
map~\eqref{eq:4DFODO} {is} controlled {by} an appropriately designed
map.

\section{Control theory for symplectic maps}
\label{sec:ctrlth}

The aim of Hamiltonian theory is to provide mathematic tools to be able to modify the dynamics of a symplectic system.  With this, it is possible to manipulate intrinsic features, e.g.~to reduce the chaotic regions in phase space or to build invariant tori.

\begin{figure}[htbp]
\begin{center}
\makebox[\textwidth][c]{
\includegraphics[width=11.5cm]{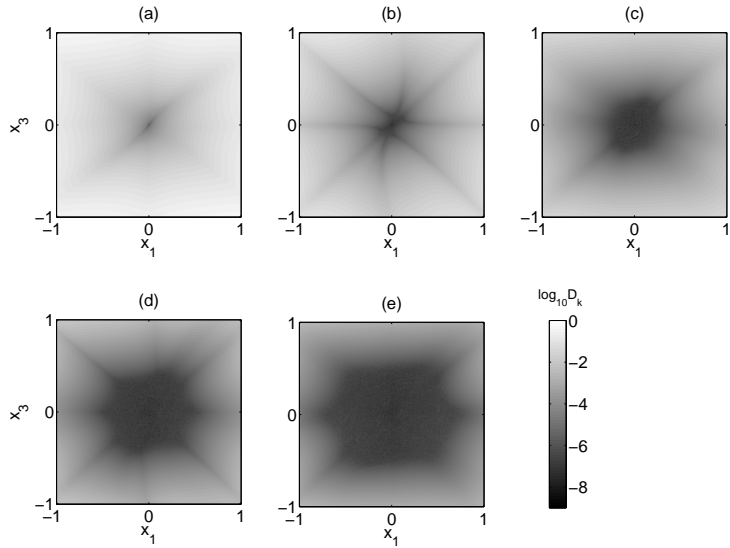}
}
\end{center}
\caption{{\em The simplicity defect of the controlled map $T_k(F_2)$
  \eqref{eq:ctrlledmapordk}}. Plot of $\log_{10}D_k$, where
  $D_k=||A_k^T{J}A_k-{J}||$ and $A_k$ is the Jacobian of the $k$-order
  controlled map $T_k(F_2)$ given by~\eqref{eq:ctrlledmapordk}, for $16000$
  uniformly distributed values {in the square} $(x_1,x_3)\in
  [-1,1]\times[-1,1]$, $x_2=x_4=0$, for (a) $k=1$, (b) $k=2$, (c) $k=3$, (d)
  $k=4$ and (e) $k=5$. {The} percentage of orbits {with
  $\log_{10}D_k < {-4}$} is $0.5\%$, $11\%$, $28\%$, $53\%$ and $73\%$ for $k=1,2,3,4$ and $5$ respectively. The gray scale corresponds to the
  value of $\log_{10}D_k$: the darker the color, the smaller the value of
  $\log_{10}D_k$ is, and hence the closer the map is to a symplectic one.}
\label{jac}
\end{figure}

In the following we will be interested in controlling a quasi-integrable
symplectic map in such {a way that it will allow us} to obtain a {new, {\em
controlled}} map {\lq\lq closer\rq\rq} to {the} integrable part {of the
original map, and thus increase the stability region around the nominal
circular orbit. This is of great importance  since chaos diffusion channels in phase space with unpredictable consequences in configuration space.
The controlled map {is expected to} have a smaller number of escaping
orbits {and} a larger region {occupied by} invariant {curves} in a
neighbourhood of the origin. For this purpose, we apply the method presented
in~\cite{CVECP2005}, with the modification that in the present case the integrable part is not
expressed in action-angle variables.  In particular, the integrable part is a rotation, so the present
theory applies {to perturbations of rotations, instead of maps close to
identity.}

Let us consider an integrable symplectic map defined through its infinitesimal
generator~\footnote{In~\cite{CVECP2005} a similar theory has been {developed}
for {a} general symplectic map devoid of an infinitesimal generator.}  $H$
\begin{equation}
\vec{x}^{\prime}=e^{\{H\}}\vec{x}\, ,
\label{eq:unpertflow}
\end{equation}
and consider the quasi-integrable map perturbation of the former
\begin{equation}
\vec{x}^{\prime}=T(\vec{x})=e^{\{H\}}e^{\{V\}}\vec{x}\, ,
\label{eq:pertflow2}
\end{equation}
where $\vec{x}\in\mathbb{R}^{2N}$ and $V$ is a perturbation, namely
$V=o(H)$. The aim of Hamiltonian control theory, is to construct a third map,
the {\em control map}, whose generator $F$ is small (it satisfies
$F=o(V)$). The {\em controlled map}
\begin{equation}
T_{ctrl}=e^{\{H\}}e^{\{V\}}e^{\{F\}}\, ,
\label{eq:controlled}
\end{equation}
will be conjugated to a map $T_*$, closer to $e^{\{H\}}$ than $T$
(see~\eqref{eq:controlconj} below). We note that the use of the
exponential of a Poisson bracket, ensures that such maps are {symplectic by
construction.}

To be more precise, let us define the unperturbed map
\begin{equation}
\label{eq:unpertmap}
\mc{A}^{-1}=e^{-\{H\}}\, ,
\end{equation}
and observe that $(1-\mc{A}^{-1})$ is not invertible, since its kernel
contains any smooth function of $H$. Thus we assume the existence of a \lq\lq
pseudo-inverse\rq\rq operator, $\mc{G}$, that should satisfy
(see~\cite{CVECP2005} for details)
\begin{equation}
\mc{G}\left(1-\mc{A}^{-1}\right)\mc{G}=\mc{G}\, .
\label{eq:GHG}
\end{equation}

At this point we can define the {\em non-resonant} and the {\em resonant}
 operators
\begin{equation}
\label{eq:NR}
\mc{N}:=\left(1-\mc{A}^{-1}\right)\mc{G}\quad\text{and}\quad\mc{R}:=1-\mc{N}\,
,
\end{equation}
which are projectors, i.e. $\mc{N}^2=\mc{N}$ and
$\mc{R}^2=\mc{R}$.

Our main theoretical result can be stated in the following theorem:
\begin{theorem}
  \label{thm:control}
Under the above hypotheses and defining $S=\mc{G}V$ we have
\begin{equation}
  \label{eq:controlconj}
  e^{\{S\}}T_{ctrl}e^{-\{S\}}=e^{\{H\}}e^{\{\mc{R}V\}}:=T_*\, ,
\end{equation}
where
\begin{equation}
  T_{ctrl}=e^{\{H\}}e^{\{V\}}e^{\{F\}}\, ,
\end{equation}
with a control term given by
\begin{equation}
  \label{eq:ctrlterm}
e^{\{F\}}=e^{-\{V\}}e^{\{\left(\mc{N-G}\right)V\}}e^{\{\mc{R}V\}}e^{\{\mc{G}V\}}\, .
\end{equation}
\end{theorem}

\begin{remark}[Warped addition]
  \label{rem:warpplus}
Let us define as in~\cite{CVECP2005} the warped addition, $\{A\}\oplus\{B\}$,
of two operators by
\begin{equation}
  \label{eq:warpadd}
  e^{\{A\}}e^{\{B\}}:=e^{\{A\}\oplus\{B\}}\, .
\end{equation}
An explicit formula can be obtained using the Baker-Campbell-Hausdorff
formula~\cite{Bourb1972}, where $\{A\}\oplus\{B\}$ is a series whose first
terms are
\begin{equation}
\{A\}\oplus\{B\}=\{A\}+\{B\}+\frac{1}{2}\left(\{A\}\{B\}-\{B\}\{A\}\right)+\dots
\, ,
\label{eq:oplus}
\end{equation}
hence the warped addition is a deformation of the usual addition between
operators.
\end{remark}

\proof
Using this warped addition, we can rewrite the controlled map into the form
\begin{equation}
\label{eq:ctrlmapoplus}
T_{ctrl}=e^{\{H\}\oplus\{V\}\oplus\{F\}}\, ,
\end{equation}
where the control term~\eqref{eq:ctrlterm} becomes
\begin{equation}
  \label{eq:ctrloplus}
  \{F\}=-\{V\}\oplus\{\left(\mc{N-G}\right)V\}\oplus\{\mc{R}V\}\oplus\{\mc{G}V\}\, .
\end{equation}

From~\eqref{eq:NR} we have:
\begin{equation}
  \label{eq:NmG}
  \mc{N-G}=-\mc{A}^{-1}\mc{G}\, .
\end{equation}
One can easily prove (see appendix A of~\cite{CVECP2005}) that
\begin{equation}
\left(-\{H\}\right)\oplus\left(-\{S\}\right)\oplus\{H\}=-\{\mc{A}^{-1}S\}\, ,
\end{equation}
hence, recalling the definition of $S$ and~\eqref{eq:NmG}, we can
rewrite~\eqref{eq:ctrloplus} as 
\begin{equation}
  \label{eq:ctrloplus2}
  \{F\}=-\{V\}\oplus (-\{H\}) \oplus (-\{S\})\oplus \{H\}
  \oplus\{\mc{R}V\}\oplus\{S\}\, .
\end{equation}
By rearranging the terms we can easily get
\begin{equation}
\{S\}\oplus\{H\}\oplus \{V\} \oplus\{F\}\oplus (-\{S\})= \{H\}
  \oplus\{\mc{R}V\}\, ,
\end{equation}
which is nothing but~\eqref{eq:controlconj} rewritten using the warped
addition.  \endproof

\begin{remark}
  \label{rem:smallness}
Let us observe that the control term is, as required, small compared to
$V$. In fact from~\eqref{eq:ctrloplus} and by using the approximated formula
for the warped addition~\eqref{eq:oplus}, we obtain
\begin{eqnarray}
  \label{eq:ctrloplus3}
  \{F\}&=&-\{V\}\oplus\{\left(\mc{N-G}\right)V\}\oplus\{\mc{R}V\}\oplus\{\mc{G}V\}\\
&=& -\{V\}+\{\mc{N}V\} -\{\mc{G}V\}+ \{\mc{R}V\}+ \{\mc{G}V\}+o(V)=o(V)\, ,
\end{eqnarray}
where we use the relation $\mc{N+R}=1$.

Under the assumption of absence of resonances, i.e. $\mc{R}V=0$, the $o(V)$
term in the control map can be explicitly computed to give (see~\ref{sec:appendix})
\begin{equation}
\label{eq:Ford2}
F=\frac{1}{2}\{V\}\mc{G}V +o(V^2)\, .
\end{equation}
\end{remark}

\section{The control term for the non-resonant  map}
\label{sec:fodoctrl}

In this section we derive the controlled map presented in section \ref{sec:ctrlth} in case of maps of the form (\ref{eq:4DFODO}).
To implement the theory we need to diagonalize the operator $\{H\}$.  For this reason we introduce complex variables
\begin{equation}
  \label{eq:complexvar}
  \zeta_1=x_2+ix_1\quad \text{and}\quad  \zeta_2=x_4+ix_3\, ,
\end{equation}
and rewrite $H$ as
\begin{equation}
  H(\zeta_1,\zeta_2)=-\frac{\omega_1}{2}|\zeta_1|^2-\frac{\omega_2}{2}|\zeta_2|^2\,
  .
\end{equation}

The Poisson bracket with $H$ now takes the form
\begin{equation}
  \{H\}=i\omega_1\left(\bar{\zeta}_1\partial_{\bar{\zeta}_1}-{\zeta}_1\partial_{{\zeta}_1}\right)+i\omega_2\left(\bar{\zeta}_2\partial_{\bar{\zeta}_2}-{\zeta}_2\partial_{{\zeta}_2}\right)\,
  .
\end{equation}
Hence, for any $\vec{n}=(n_1,n_2)\in\mathbb{N}^2$ and
$\vec{m}=(m_1,m_2)\in\mathbb{N}^2$ we obtain
\begin{equation}
  \{H\}\zeta^{\vec{n}}
  \bar{\zeta}^{\vec{m}}=i\left(\omega_1m_1-\omega_1n_1+\omega_2m_2-\omega_2n_2\right)\zeta^{\vec{n}}
  \bar{\zeta}^{\vec{m}}=i\vec{\omega}\cdot(\vec{m}-\vec{n})\zeta^{\vec{n}}
  \bar{\zeta}^{\vec{m}}.
\end{equation}
Here we introduced the vector $\vec{\omega}=(\omega_1,\omega_2)$ and use
the compact notation $\zeta^{\vec{n}}=\zeta_1^{n_1}\zeta_2^{n_2}$ for the
complex vector ${\zeta}=(\zeta_1,\zeta_2)$. The operator $\{H\}$ is diagonal
in these variables and thus map~\eqref{eq:unpertmap} is straightforwardly
obtained as
\begin{equation}
  \mc{A}^{-1}\zeta^{\vec{n}}\bar{\zeta}^{\vec{m}}=e^{-\{H\}}\zeta^{\vec{n}}\bar{\zeta}^{\vec{m}}=e^{-i\vec{\omega}\cdot(\vec{m}-\vec{n})}\zeta^{\vec{n}}\bar{\zeta}^{\vec{m}}\, .
\end{equation}

Once we have this map, we can compute the operators $\mc{G}$, $\mc{N}$ and
$\mc{R}$. For all $\vec{n}$ and
$\vec{m}\in\mathbb{N}^2\setminus \{0\}$, such that $\vec{n}\neq\vec{m}$ and 
\begin{equation}
  \label{eq:nonres}
  \vec{\omega}\cdot(\vec{m}-\vec{n})\neq 2k\pi\quad \forall k\in\mathbb{Z}\, 
\end{equation}
(which defines the {\em
  non-resonance condition}), {we get}
\begin{equation}
  \label{eq:Gcmplx}
  \mc{G}\zeta^{\vec{n}}\bar{\zeta}^{\vec{m}}=\frac{1}{1-e^{-i\vec{\omega}\cdot(\vec{m}-\vec{n})}}\mc{N}\zeta^{\vec{n}}\bar{\zeta}^{\vec{m}}\, ,
\end{equation}
with
\begin{equation}
  \label{eq:NRcmplx}
  \mc{N}\zeta^{\vec{n}}\bar{\zeta}^{\vec{m}}=\begin{cases}\zeta^{\vec{n}}\bar{\zeta}^{\vec{m}}
    & \text{if $\vec{\omega}\cdot(\vec{m}-\vec{n})\neq 2k\pi\quad\forall k\in\mathbb{Z}$}\\
0 & \text{otherwise}
\end{cases}\, .
\end{equation}

In the rest of the paper, we will assume the above non-resonant
condition~\eqref{eq:nonres} to hold for the considered values of $q_x$ and
$q_y$.  Let us remark that this is not a limitation of the actual theory, but
just a working assumption. We could equivalently have chosen to work in the
resonant regime, using a different control term suitable for resonant
dynamics.

The explicit computations are as follows : first we need to express $V$
in terms of complex variables; next we compute $S=\mc{G}V$, and transform back to the original
variables. Finally we compute the
exponential $e^{\{\mc{G}V\}}$. Actually, even if $\mc{G}V$ is a polynomial of
degree three in the variables $x_1,x_2,x_3,x_4$ the map $e^{\{\mc{G}V\}}$ is
given by an infinite series. The terms of this series can be sequentially computed, but the
degree of complexity (i.e.~the number of involved terms) increase very
fast. To simplify the computations we decided to use the {\em approximated} generator of the
control map \eqref{eq:Ford2} already truncated at order 2
\begin{equation}
  F_2=\frac{1}{2}\{V\}\mc{G}V\, .
\end{equation}
{We note that} $F_2$ is composed by about twenty terms. A detailed discussion
of the whole procedure needed to {obtain} $F_2$, as well as its explicit
formula, are presented in~\ref{sec:appendix}.

We {now face} another difficulty, namely {the computation of} the control map
from the generator $F_2$. This is equivalent to perform the sum 
\begin{equation}
  e^{\{F_2\}}=1+\{F_2\}+\frac{1}{2}\{F_2\}^2 +\dots \, ,
\label{eq:sum}
\end{equation}
whose complexity once again grows very fast. We thus introduce a second
approximation {to our construction}, by computing only a finite number of
terms in {the above} sum{. So, we define} a {\em truncated control map of
order} $k$
\begin{equation}
  \label{eq:ctrlmapordk}
  C_k(F_2)=\sum_{l=0}^{k}\frac{\{F_2\}^l}{l!}\, ,
\end{equation}
and a {\em truncated controlled map of order} $k$
\begin{equation}
  \label{eq:ctrlledmapordk}
  T_k(F_2)=e^{\{H\}}e^{\{V\}} C_k(F_2)=T\,C_k(F_2)\, .
\end{equation}

Since the exact control map $e^{\{F_2\}}$ given by~\eqref{eq:sum}} is symplectic, the
controlled map~\eqref{eq:controlled} will also be symplectic. On the other
hand, we cannot expect the $k$-th order control map $ C_k(F_2)$ to be
symplectic. We know that a map is symplectic if its Jacobian matrix $A$
verifies (in its definition domain) the equality
\begin{equation}
  A^{\mathrm{T}}JA-J=0\, ,
\end{equation}
Thus, in order to check the {\em symplecticity defect} of $T_k(F_2)$ {we
compute} the norm {$D_k$} of matrix $A_k^{\mathrm{T}}JA_k-J$, where $A_k$ is
the Jacobian of {$T_k(F_2)$ given by~\eqref{eq:ctrlledmapordk}.}  The results
are presented in Fig.~\ref{jac} for orders $k=1$ up to $k=5$, in {the region}
$(x_1,x_3)\in [-1,1]\times[-1,1]$, $x_2=x_4=0$. The results indicate that
$T_k(F_2)$ is a good approximation of a symplectic map for $k\geq 4$, {because
we get $D_k\lesssim 10^{-4}$} for a large {portion ($\gtrsim 53$ \%)} of
variables values. {We note that in the central region of the truncated
controlled map, where the actual physical process of beam's evolution occurs,
the symplectic character of the map is established even better since there $D_k
\lesssim 10^{-8}$.} As expected, the larger the order $k$, the closer to
symplecticity the approximation {is}.

\begin{figure}[htbp]
\begin{center}
\makebox[\textwidth][c]{
\includegraphics[width=16.5cm]{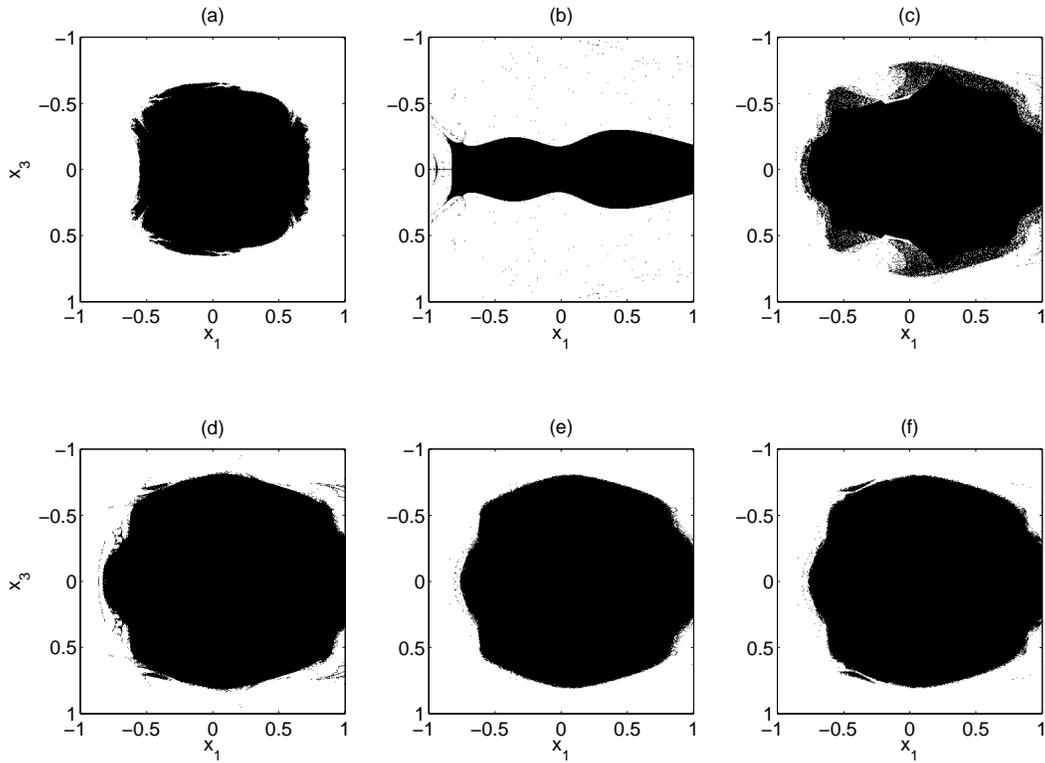}
}
\end{center}
\caption{{\em Non-escaping regions of controlled map $T_k(F_2)$ as a function
    of the truncation order $k$}. $16000$ uniformly distributed initial
    conditions in the square $(x_1,x_3)\in[-1,1]\times [-1,1]$,
    $x^{(0)}_2=x^{(0)}_4=0$ are iterated up to $n=10^5$ using (a) the uncontrolled
    map~\eqref{eq:4DFODO} and (b)--(f) the $k=1$ to $k=5$ order controlled map
    $T_k(F_2)$~\eqref{eq:ctrlledmapordk}, respectively. Initial conditions
    corresponding to non-escaping orbits up to $n=10^5$ are coloured in black,
    while escaping orbits are colored in white.}
\label{compare}
\end{figure}

\begin{figure}[htbp]
\begin{center}
\includegraphics[height=7cm]{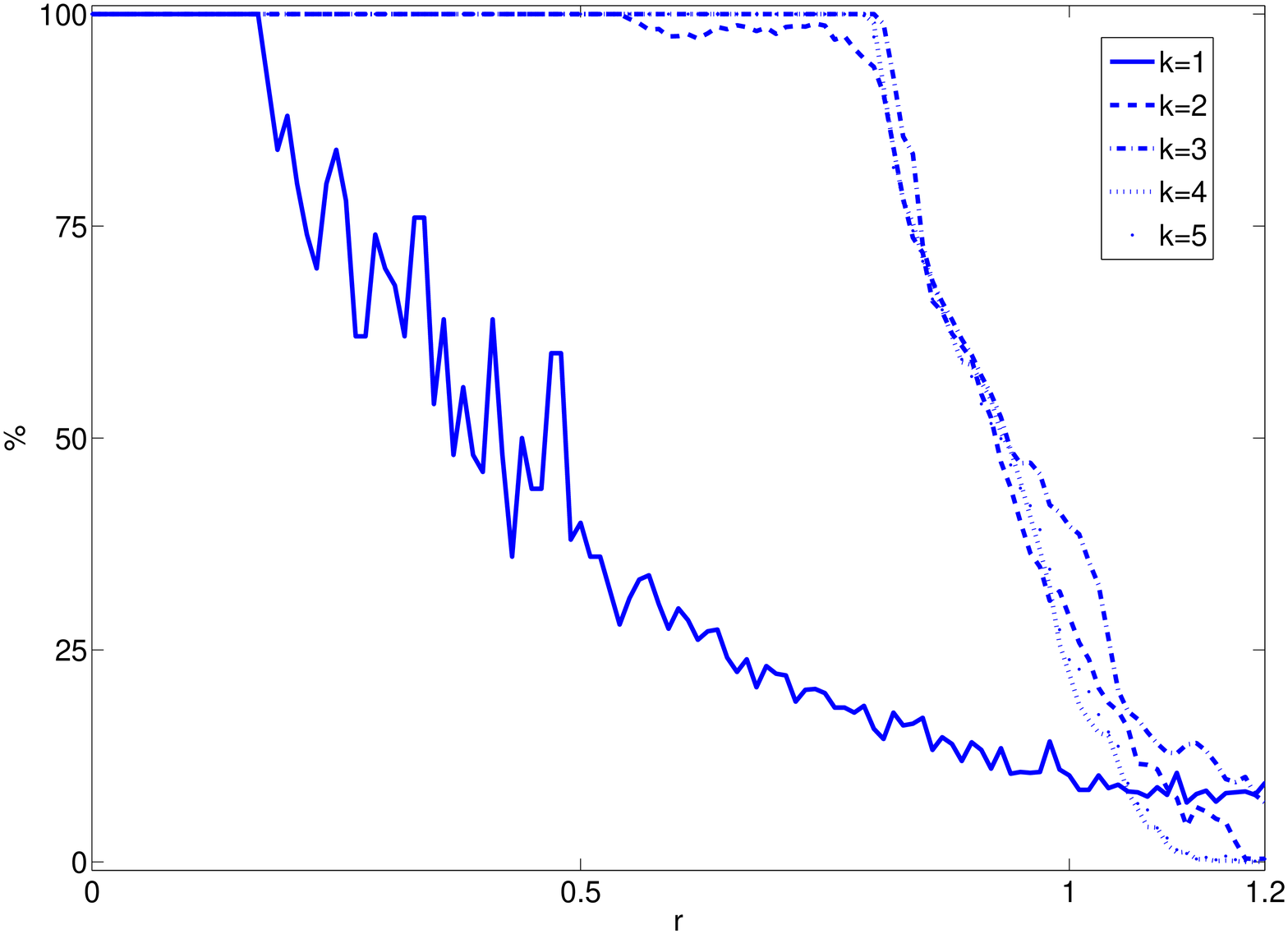}
\end{center}
\caption{{\em Percentages of non-escaping orbits for the controlled map
 $T_k(F_2)$~\eqref{eq:ctrlledmapordk} as a function of the distance from the
 origin in the physical space $(x_1,x_3)$}. We iterate initial conditions in a
 circle of radius $r$ centred at the origin of plane $(x_1,x_3)$, with
 $x_2^{(0)}=x_4^{(0)}=0$, and compute the percentages of non-escaping orbits during
 $n=10^5$ iterations, for the controlled map $T_k(F_2)$ with $k=1,2,3,4,5$, as
 a function of $r$. }
\label{controlcompare}
\end{figure}

{The main objective of the addition of a control term is to increase the size
of the stability region around the central periodic orbit. This increase leads
to decrease the number of escaping orbits\footnote{An orbit $\left(x^{(k)}_1,x^{(k)}_2,x^{(k)}_3,x^{(k)}_4\right)_{0\leq k\leq N}$ is defined as non--escaping if for all $k\leq N\,\,{x^{(k)}_1}^2+{x^{(k)}_2}^2+{x^{(k)}_3}^2+{x^{(k)}_4}^2\leq R^2$ for a certain $R$ (in the simulations we used $R^2=10$ and $N=10^4,10^5$) and escaping otherwise}, as we can see from the results presented in Fig.~\ref{compare}, where we plot
in black the initial conditions on the square $(x_1,x_3)\in
[-1,1]\times[-1,1]$, $x_2=x_4=0$, giving rise to orbits that do not escape up
to $10^5$ iterations of the map. In particular, we consider in
Fig.~\ref{compare}(a) the original uncontrolled map~\eqref{eq:4DFODO}, and in
Figs.~\ref{compare}(b) to~\ref{compare}(d), the $k$ order controlled map
$T_k(F_2)$ for $k=1$ to $k=5$ respectively.} One can easily see that the
region of non-escaping orbits for the original map is smaller than the one of
the controlled maps. This observation can be quantified by considering initial
conditions inside a circle centered at the origin of each panel of
Fig.~\ref{compare} (which represent the actual physical plane since the
initial momenta are $x^{(0)}_2=x^{(k)}_4=0$) with radius $r^2={x^{(0)}_1}^2+{x^{(0)}_3}^2$,
and evaluate the number of escaping and non-escaping orbits as a function of
the circle radius for $T_k(F_2)$ with $k=1$ up to $k=5$.  Results reported in
Fig.~\ref{controlcompare} support the previous claim, by clearly showing that
controlled maps of orders $3$, $4$ and $5$ behave very similarly and lead to
an increase of the non-escaping region.  Let us note that the behavior of the
controlled maps of orders $k=1$ (Fig.~\ref{compare}(b)) and $k=2$
(Fig.~\ref{compare}(c)) is somewhat misleading if it is not analyzed together
with the information from the symplecticity defect (see Figs.~\ref{jac}(a) and
(b) respectively). In fact, these maps are strongly dissipative and produce a
strong shift of orbits towards the origin, preventing them from escaping. This
dissipation effect is not physical, as it is not observed in real
accelerators, and therefore we do not discuss further the $k=1$ and $k=2$
controlled maps.

{From the results of Figs.~\ref{compare} and~\ref{controlcompare} we see that
the addition of even the lower order ($k=3$) control term, having an
acceptable symplecticity defect, increases drastically the size of the region
of non-escaping orbits around the central periodic orbit.  A further increase
of the order of the control term results to less significant increment of this
region, while the computational effort for constructing the controlled map
increases considerably. In fact, $T_1(F_2)$ contains around $100$ elementary
terms, i.e.~monomials in $x_1\dots x_4$, while this number is almost doubled for each order, so that $T_5(F_2)$
contains around $2000$ terms. Also the CPU time needed to evolve the
orbits increases with the order. For example, while the integration of one
orbit using $T_1(F_2)$ takes about $1.4$ times the CPU time needed to
integrate the original map (\ref{eq:4DFODO}), the use of $T_5(F_2)$ needs
almost $21.5$ times more.}

{Thus we conclude that the $T_4(F_2)$ controlled map, which can be considered
quite accurately to be symplectic, is sufficient to get significant increment
of the percentage of non-escaping orbits, without paying an extreme
computational cost}.

\section{The SALI method}
\label{sec:salimeth}
The Smaller Alignment Index (SALI)~\cite{S01} has been proved to be an
efficiently simple method to determine the regular or chaotic nature of
orbits in conservative dynamical systems. Thanks to its properties, it has already been successfully distinguished between regular and chaotic motion
both, in symplectic maps and Hamiltonian flows
\cite{SABV03,SABV04,SESS04,PBS04,MA05}.

For the sake of completeness, let us briefly recall the definition of the SALI
and its behavior for regular and chaotic orbits, restricting our attention to
$2N$-dimensional symplectic maps. The interested reader can consult~\cite{S01}
for a more detailed description.  To compute the SALI of a
given orbit of such maps, one has to follow the time evolution of the orbit
itself and also of two linearly independent unitary deviation vectors
$\hat{v}_{1}^{(0)},\hat{v}_{2}^{(0)}$. The evolution of an orbit of a map $T$ is
described by the discrete-time equations of the map
\begin{equation}
\label{map_eq:1}
    \vec{x}^{(n+1)}=T(\vec{x}^{(n)})\, ,
\end{equation}
where $\vec{x}^{(n)}=(x_{1}^{(n)},x_{2}^{(n)},...,x_{2N}^{(n)})^{\rm T}$,
represents the orbit's coordinates at the $n$-th iteration. The deviation
vectors $\vec{v}_{1}^{(n)},\vec{v}_{2}^{(n)}$ at time $n$ are given by the
tangent map
\begin{equation}
\label{tan_mao:1}
    \vec{v}_i^{(n+1)}=A(\vec{x}^{(n)})\cdot \vec{v}_i^{(n)} \quad i=1,2\, ,
\end{equation}
where $A$ denotes the Jacobian matrix {of  map}~\eqref{map_eq:1}, evaluated
at the points of the orbit under study. {Then,} according to~\cite{S01} the
SALI for the given orbit is defined as
\begin{equation}\label{eq:SALI:2}
\mbox{SALI}(n)=min
\left\{\left\|\hat{v}_{1}^{(n)}+\hat{v}_{2}^{(n)}\right\|,\left\|\hat{v}_{1}^{(n)}-\hat{v}_{2}^{(n)}\right\|\right\}\,
  ,
\end{equation}
where $\| \cdot \|$ denotes the usual Euclidean norm and
$\hat{v}_{i}=\frac{\vec{v}_{i}}{\| \vec{v}_{i}\|}$, $i=1,2$ are unitary
normalised vectors.

In the case of chaotic orbits, the deviation vectors $\hat{v}_1$, $\hat{v}_2$
eventually become aligned in the direction defined by the maximal Lyapunov
characteristic exponent (LCE), and SALI$(n)$ falls exponentially to zero. An
analytical study of SALI's behavior for chaotic orbits was carried out
in~\cite{SABV04} where it was shown that
\begin{equation}
\mbox{SALI}(n) \propto e^{-(\sigma_1-\sigma_2)n},
\label{eq:expsali}
\end{equation}
with $\sigma_1$, $\sigma_2$ being the two largest LCEs.

{On the other hand, in the case of regular motion} the orbit lies on a torus
and the vectors $\hat{v}_1$, $\hat{v}_2$ eventually fall on its tangent space,
following a $n^{-1}$ time evolution, having in general different directions.
This behavior is due to the fact that for regular orbits, the norm of a
deviation vector increases linearly in time. Thus, the normalization procedure
brings about a decrease of the magnitude of the coordinates perpendicular to
the torus, at a rate proportional to $n^{-1}$, and so $\hat{v}_1$, $\hat{v}_2$
eventually fall on the tangent space of the torus. In this case, the SALI
oscillates about non-zero values (for more details see~\cite{SABV03}).

The simplicity of SALI's definition, its completely different behavior for
regular and chaotic orbits and its rapid convergence to zero in the case of
chaotic motion are the main advantages that make SALI an ideal chaos detection
tool. Recently a generalization of the SALI, the so-called Generalized
Alignment Index (GALI) has been introduced~\cite{SBA07,SBA08}, which uses
information of more than two deviation vectors from the reference orbit. Since
the advantages of GALI over SALI become relevant in the case of
multi-dimensional systems, in the present paper we apply the SALI method for
the dynamical study of the 4D map (\ref{eq:4DFODO}).

\section{Dynamics of the controlled  map}
\label{sec:results}

\begin{figure}[htbp]
\begin{center}
\makebox[\textwidth][c]{
\includegraphics[width=15cm]{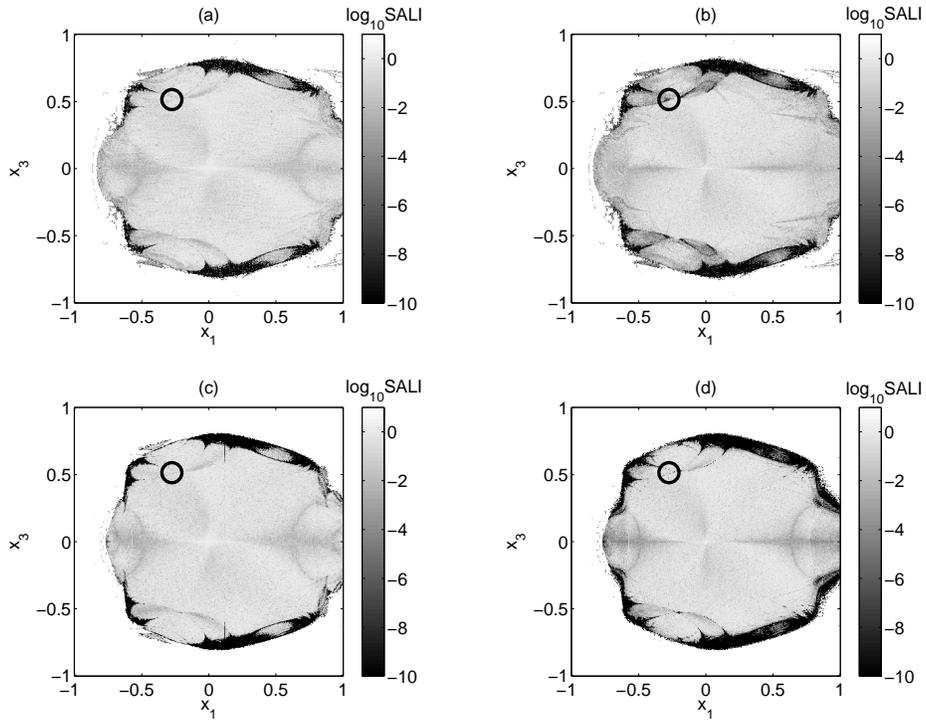}
}
\end{center}
\caption{{\em Stability analysis in the $(x_1,x_3)$ plane}.  $16000$ uniformly
  distributed initial conditions in the square $(x_1,x_3)\in[-1,1]\times
  [-1,1]$, $x_2^{(0)}=x_4^{(0)}=0$ are integrated using the $T_3(F_2)$ ((a) and
  (b)), and the $T_4(F_2)$ controlled map ((c) and (d)), up to $n=10^4$ ((a)
  and (c)) and $n=10^5$ iterations ((b) and (d)). The gray scale represents
  the value of $\log_{10}$SALI for each orbit at the end of the integration
  time. The lighter the color the more stable is the orbit, while white color
  denotes that an orbit escaped before the total number of iterations was
  reached. The black circle indicate the initial condition of the orbit
  studied in Fig.~\ref{orbite}.}
\label{fig5c}
\end{figure}

As already mentioned, the goal of constructing the controlled map
$T_{ctrl}=T\, e^{\{F\}}$ is to increase the percentage of {regular} orbits
up to a given (large) number of iterations, or equivalently increase {the size
of} the stability region around the nominal circular trajectory {(i.e.~the
DA)}. Because the presence of chaotic regions can induce a large drift in the
phase space, that eventually could lead to the escape of orbits, the
achievement of a larger DA can be qualitatively inspected by checking via the
SALI method the regular or chaotic nature of orbits in a neighborhood of the
origin (see Fig.~\ref{fig5c}). We note that we define an orbit to be chaotic
whenever SALI$(t) < 10^{-8}$, and regular for the contrary.

In Fig.~\ref{fig5c} (which should be compared with Fig.5 of \cite{BS2006}) we
observe a strong enlargement of the region of regular orbits. This region is
characterized by large SALI values. In particular, for $10^4$ iterations of
the $T_4(F_2)$ map, $54\%$ of the considered orbits are regular, while for
the uncontrolled map this percentage reduces to $33\%$. This improvement can
be also confirmed by visual inspection of Fig.~\ref{compare}, where the regions
of non-escaping orbits are shown for different orders of the controlled map
(\ref{eq:ctrlledmapordk}).

\begin{figure}[htbp]
\begin{center}
\includegraphics[width=10cm]{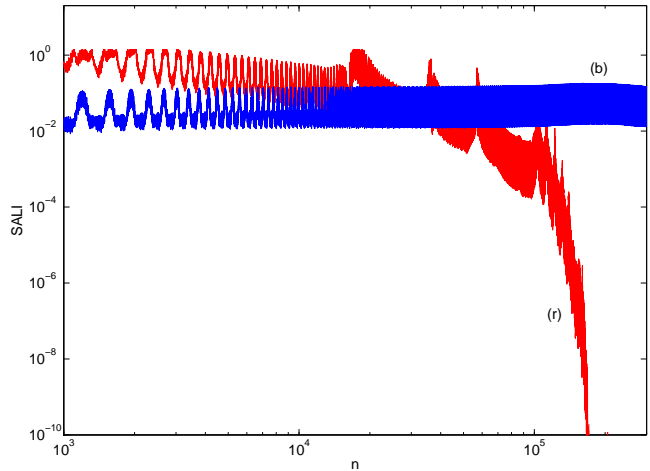}
\end{center}
\caption{{\em (Color online) Dynamics of two orbits with the same initial
  conditions for the $3^{rd}$ and $4^{th}$ order controlled maps}. Time
  evolution of the SALI for the orbit with initial conditions
  $\vec{x}^{(0)}=(-0.50,0,-0.65,0)^T$ (see Fig.~\ref{fig5c}), using the
  $T_3(F_2)$ [(r) red curve] and the $T_4(F_2)$ controlled map [(b) blue
  curve].}
\label{orbite}
\end{figure}

In Figs.~\ref{fig5c}(a) and (b) we see that there exist orbits of the
$T_3(F_2)$ map, which are characterized as regular up to $n = 10^4$
iterations, while they show their chaotic character once they are iterated up
to $n=10^{5}$. Such orbits correspond to the dark regions marked by a black
circle in Fig.~\ref{fig5c}(b) (for comparison this circle is also plotted in
all panels of Fig.~\ref{fig5c}). This discrepancy is absent for the $T_4(F_2)$
map, which shows almost the same geometrical shape for the
non-escaping region when we pass from $10^4$ to $10^5$ iterations. In order to
better understand this behavior we followed the evolution of a single orbit
with initial condition $\vec{x}^{(0)}=(-0.50,0,-0.65,0)^T$ --inside the black
circle in Fig.~\ref{fig5c}-- for both the $T_3(F_2)$ and the $T_4(F_2)$
controlled maps, computing the corresponding SALI values up to $2\times 10^5$
iterations. The results are reported in Fig.~\ref{orbite} and clearly show
that the orbit behaves regularly up to $n \approx 10^5$ iterations of the
$T_3(F_2)$ map, since its SALI values are different from zero, but later on
a sudden decrease of SALI to zero denotes the chaotic character of the
orbit. This behavior clearly implies this is a slightly chaotic, sticky orbit,
which remains close to a torus for long time intervals ($n \approx 10^5$),
while later on it enters a chaotic region of the phase space. It is
interesting to note that iterating the same initial condition by the
$T_4(F_2)$ map we get a regular behavior at least up to $n=2\times 10^5$.

\begin{figure}[htbp]
 \begin{center}
 \makebox[\textwidth][c]{
 \includegraphics[width=17.5cm]{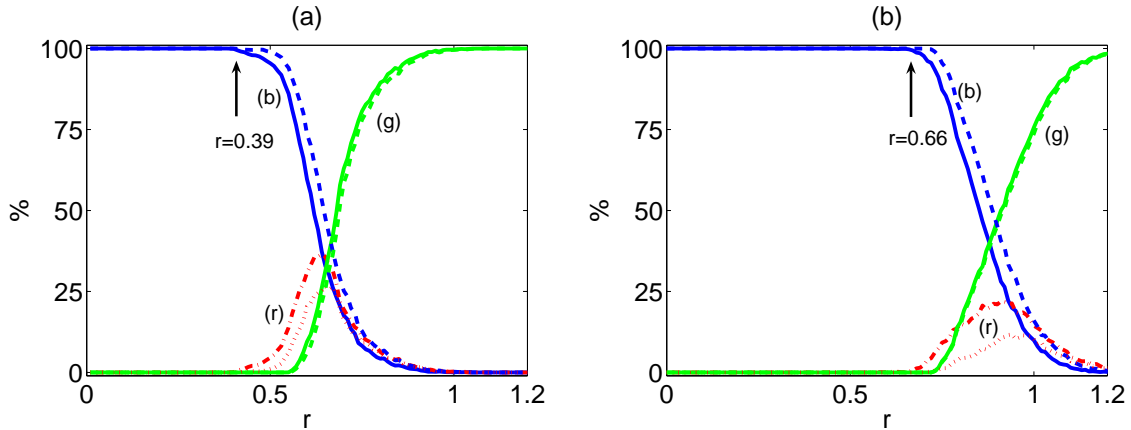}
}
 \end{center}
\caption{{\em (Color online) Dynamical aperture of the (a) original map
(\ref{eq:4DFODO}) and (b) the $T_4(F_2)$ controlled map
(\ref{eq:ctrlledmapordk})}. The percentages of regular [(b) blue curves],
escaping [(g) green curves] and chaotic [(r) red curves] orbits after
$n=10^4$ (dashed curves) and $n=10^5$ iterations (solid curves) for initial
conditions in a 4D sphere centred at the origin $x_1=x_2=x_3=x_4=0$, as a
function of the sphere radius $r$. Each point corresponds the average value
over $5000$ initial conditions. The largest radius at which the percentage of
regular orbits is still $100\%$, is marked by an arrow in each panel.}
 \label{sphere}
 \end{figure}

In order to provide additional numerical evidence of the effectiveness of the
controlled map (\ref{eq:ctrlledmapordk}) in increasing the DA, we consider
initial conditions inside a 4D sphere centered at the origin
$x_1=x_2=x_3=x_4=0$ of the map, with radius,
$r^2={{x_1^{(0)}}^2+{x_2^{(0)}}^2+{x_3^{(0)}}^2+{x_4^{(0)}}^2}$. We compute the number of
regular, escaping and chaotic orbits as a function of the sphere radius.  The
corresponding results are reported in Fig.~\ref{sphere}(b), while in
Fig.~\ref{sphere}(a) we reproduce Fig.~6 of \cite{BS2006} for comparison. From
this figure we observe a strong increase of the DA, since the largest sphere
containing $100\%$ regular orbits has a radius $r\approx0.66$, while this
radius was $r\approx0.39$ for the original uncontrolled map. We also observe
that increasing the total number of iterations from $10^4$ to $10^5$
 (dashed and solid lines in Fig.~\ref{sphere} respectively) increases the
percentage of chaotic orbits, but the radius of the 4D sphere
containing only regular orbits does not change significantly.

\section{Conclusions}
\label{sec:ccl}

In this paper we considered a simple model of a ring particle accelerator
with sextupole nonlinearity that can
be described by a symplectic map. In the framework of Hamiltonian control
theory, we were able to control the dynamics of the original system, by
providing a suitable control map, resulting in a small \lq\lq
perturbation\rq\rq of the initial map. This control map has been constructed
with the aim of DA enlargement of the particle accelerator, and thus improving
the beam's life-time and the accelerator's performance.

In particular, the theoretical framework we developed allows
a $1$-parameter family of approximated controlled maps. We performed several
numerical simulations in order to choose \lq\lq the best\rq\rq approximated
controlled map $T_k(F_2)$ (\ref{eq:ctrlledmapordk}), taking into account the
complexity of the map, i.e.~the number of terms by which it is composed, the
CPU time needed to perform the numerical iteration of orbits, and the accuracy
of the results in terms of the symplectic character of the map. We find
that the $4^{th}$ order controlled map $T_4(F_2)$ is an optimal choice for the
controlled system.

Using this controlled map we succeeded in achieving our initially set
 goal, since the $T_4(F_2)$ map exhibits a DA with a radius more than $1.7$
 times larger than the one for the original map (see Fig.~\ref{sphere}).

\appendix
\section{Computation of the control term.}
\label{sec:appendix}

The aim of this section is to introduce further details for the construction
of the control term and of the controlled map, and to provide explicit
formulas for the interested reader.

\subsection{Notations}

Let us first introduce some notations and recall some useful relations.
\begin{itemize}
\item {\bf Lie brackets and operators.} Let $\mathcal{X}$ be the vector space
 of $C^{\infty}$ real or complex functions of $2N$ variables $(p,q)$.  For any
 $F,\,G\, \in \mathcal{X}$, the Lie bracket is given by
\begin{equation}
\{F,G\}:=\sum_{i=1}^{N}\left[\partial_{p_i}F\partial_{q_i}G-\partial_{q_i}F
\partial_{p_i}G\right]\, ,
\end{equation}
where $\partial_{x_i}f\equiv \frac{\partial f}{\partial x_i}$ denotes the
partial derivative with respect to the variable $x_i$.

Using the above definition, we can define a linear operator, induced by an
element $F$ of $\mathcal{X}$, acting on $\mathcal{X}$
\begin{equation}
\label{eq:crochet}
\begin{array}{rccl}
\{F\}: & \mathcal{X} & \rightarrow & \mathcal{X} \\
       &   G         & \mapsto     & \{F\}G:=\{F,G\}
\end{array}
\end{equation}
This operator is \textit{linear}, \textit{antisymmetric} and verifies the
\textit{Jacobi identity} 
\begin{equation}
\label{jacobi}
\forall F,G\in\mathcal{X}\quad\{\{F\}G\}=\{F\}\{G\}-\{G\}\{F\}
\end{equation}
\item {\bf Exponential.} We define the exponential of such an operator
 $\{F\}$, by 
\begin{equation}
\label{exponential}
e^{\{F\}}:=\sum_{k=0}^{\infty}\frac{\{F\}^k}{k!}\, ,
\end{equation}
which is also an operator acting on $\mathcal{X}$.  The power of an operator
is the composition : $\{F\}^kG=\{F\}^{k-1}\left(\{F\}G\right)$.

We observe that in the case of the Hamiltonian function $H$, the exponential
provides the flow, namely $e^{\{H\}}x_0=x(t)$, of the Hamilton equations
\begin{equation}
  \begin{cases}
    \dot{p}&=-\partial_q H\\
    \dot{q}&=\partial_p H\, .
  \end{cases}
\end{equation}
\item {\bf Vector Field.} The action of the above defined operators, can be
  extended to vector fields \lq\lq component by component\rq\rq
\begin{equation}
\forall\,F,G,H\in \mathcal{X}\quad\{F\}
\left(
\begin{array}{c}
G \\
H
\end{array}
\right)
:=
\left(
\begin{array}{c}
\{F\}G \\
\{F\}H
\end{array}
\right).
\end{equation}
\end{itemize}

\subsection{Mappings as time-$1$ flows.}
We show now that map (\ref{eq:4DFODO}) can be seen as the time-$1$ flow of a given
Hamiltonian system. More precisely we show that
\begin{eqnarray}
\label{mappings}
T
\left(
\begin{array}{c}
x_1 \\
x_2 \\
x_3 \\
x_4
\end{array}
\right) & = &
\left(
\begin{array}{cccc}
\cos(\omega_1) & -\sin(\omega_1) & 0 & 0 \\
\sin(\omega_1) & \cos(\omega_1) & 0 & 0 \\
0&0&\cos(\omega_2) & -\sin(\omega_2) \\
0&0&\sin(\omega_2) & \cos(\omega_2)
\end{array}
\right)
\left(
\begin{array}{c}
x_1 \\
x_2+x_1^2-x_3^2 \\
x_3 \\
x_4-2x_1x_3 
\end{array}
\right) \nonumber \\
&=&
e^{\{H\}}e^{\{V\}}\left(
\begin{array}{c}
x_1 \\
x_2 \\
x_3 \\
x_4 
\end{array}
\right)
\end{eqnarray}
where
\begin{equation}
H(x_1,x_2,x_3,x_4)=-\omega_1\frac{x_1^2+x_2^2}{2}-\omega_2\frac{x_3^2+x_4^2}{2}
\end{equation}
and
\begin{equation}
\label{V}
V(x_1,x_2,x_3,x_4)=-\frac{x_1^3}{3}+x_1x_3^2\, .
\end{equation}

Let us observe that $H$ is the sum of two non-interacting harmonic oscillators
with frequencies $\omega_1$ and $\omega_2$, hence its dynamics is explicitly
given by
\begin{eqnarray}
\begin{cases}
x_1(t)&=A\cos(\omega_1t)-B\sin(\omega_1t)\\
x_2(t)&=B\cos(\omega_1t)+A\sin(\omega_1t)\\
x_3(t)&=C\cos(\omega_2t)-D\sin(\omega_2t)\\
x_4(t)&=D\cos(\omega_2t)+C\sin(\omega_2t)\, .
\end{cases}
\end{eqnarray}

By definition $\vec{y}=e^{\{H\}}\vec{x}$ is the solution at time $1$ with
initial condition $\vec{x}=(x_1,x_2,x_3,x_4)^T$, hence we obtain
\begin{eqnarray}
\begin{cases}
y_1&=\cos(\omega_1)x_1-\sin(\omega_1)x_2\\
y_2&=\sin(\omega_1)x_1+\cos(\omega_1)x_2\\
y_3&=\cos(\omega_2)x_3-\sin(\omega_2)x_4\\
y_4&=\sin(\omega_2)x_3+\cos(\omega_2)x_4\, ,
\end{cases}
\end{eqnarray}
that is 
\begin{equation}
\label{flowH}
e^{\{H\}}\left(
\begin{array}{c}
x_1 \\
x_2 \\
x_3 \\
x_4 
\end{array}
\right)=
\left(
\begin{array}{cccc}
\cos(\omega_1) & -\sin(\omega_1) & 0 & 0 \\
\sin(\omega_1) & \cos(\omega_1) & 0 & 0 \\
0&0&\cos(\omega_2) & -\sin(\omega_2) \\
0&0&\sin(\omega_2) & \cos(\omega_2)
\end{array}
\right)
\left(
\begin{array}{c}
x_1 \\
x_2 \\
x_3 \\
x_4 
\end{array}
\right)\, .
\end{equation}

From~\eqref{V} and the definition~\eqref{eq:crochet} we easily get
\begin{eqnarray}
\{V\}&:=&\partial_{x_2}V\partial_{x_1}-\partial_{x_1}V\partial_{x_2}+\partial_{x_4}V\partial_{x_3}-\partial_{x_3}V\partial_{x_4}\\
&=&(x_1^2-x_3^2)\partial_{x_2}-2x_1x_3\partial_{x_4}.\nonumber
\end{eqnarray}
This means that once applied to a vector $\vec{x}$ only the second and fourth
components of $\{V\}\vec{x}$ will be non-zero and moreover they only depend on
the first and third components of $\vec{x}$, hence $\{V\}^2\vec{x}=\vec{0}$.
We can thus conclude that $\forall\,k\geq 2$ and $\forall
\vec{x}\in\mathbb{R}^4$, we get $\{V\}^k\vec{x}=\vec{0}$.  Finally using the
definition (\ref{exponential}) we obtain
\begin{equation}
e^{\{V\}}\vec{x}=\sum_{k=0}^{\infty}\frac{\{V\}^k}{k!}\vec{x}=I\vec{x}+\{V\}\vec{x}=
\left(
\begin{array}{c}
x_1 \\
x_2 +x_1^2-x_3^2\\
x_3 \\
x_4 -2x_1x_3
\end{array}
\right)
\end{equation}

\subsection{Computation of the generator $F$ under the assumption
  $\mc{R}V\equiv 0$.}

Let us recall that the composition of maps expressed by exponential defines
the warped addition
\begin{equation}
e^{\{A\}}e^{\{B\}}:=e^{\{A\}\oplus\{B\}}\, ,
\end{equation}
whose first terms are
\begin{equation}
\{A\}\oplus\{B\}=\{A\}+\{B\}+\frac{1}{2}(\{A\}\{B\}-\{B\}\{A\})+\dots
\end{equation}

Using the warped addition with equation \eqref{eq:ctrlterm} of
Theorem~\ref{thm:control}, we obtain
\begin{equation}
e^{\{F\}}=e^{-\{V\}}e^{\{(\mc{N}-\mc{G})V\}}e^{\{\mc{R}V\}}e^{\{\mc{\mc{G}V}\}}
=e^{-\{V\}\oplus\{(\mc{N}-\mc{G})V\}\oplus\{\mc{R}V\}\oplus\{\mc{\mc{G}V}\}}\, ,
\end{equation}
and thus
\begin{eqnarray}
 \{F\}&=&-\{V\}\oplus\{(\mc{N}-\mc{G})V\}\oplus\{\mc{R}V\}\oplus\{\mc{\mc{G}V}\}=-\{V\}\oplus\{(1-\mc{G})V\}\oplus\{\mc{G}V\}+o(V^2)\nonumber\\
&=&\left[\frac{1}{2}\underbrace{\left(\{V\}\{\mc{G}V\}-\{\mc{G}V\}\{V\}\right)}_{\stackrel{(\ref{jacobi})}{=}\{\{V\}\mc{G}V\}}-\{\mc{G}V\}\right]\oplus\{\mc{G}V\}+o(V^2)\nonumber\\
&=&\frac{1}{2}\{\{V\}\mc{G}V\}+\frac{1}{4}\underbrace{\{\{\{V\}\mc{G}V\}\mc{G}V\}}_{=o(V^2)}+o(V^2)=\frac{1}{2}\{\{V\}\mc{G}V\}+o(V^2)\,
,
\end{eqnarray}
where we explicitly used the assumption $\mc{R}V=0$ to remove the third term
on the the right hand side on the first equation and hence to write
$\mc{N}V=V$. We are thus able to define the non-resonant control term, up to
order $V^2$, to be
\begin{equation}
\label{eq:f2}
F_2=\frac{1}{2}\{V\}\mc{G}V=\frac{1}{2}\{V,\mc{G}V\}\, .
\end{equation}

\subsection{The operator $\mathcal{G}$}
To get the explicit formula for $F_2$ we need to compute the expression of
$\mc{G}$.
>From definition~\eqref{eq:GHG} the operator $\mc{G}$ should satisfy
\begin{equation}
\label{propG}
\mathcal{G}\left(1-e^{-\{H\}}\right)\mathcal{G}=\mathcal{G}\, .
\end{equation}
To construct it, it will be more convenient to use complex variables
\begin{equation}
\label{complexCoor}
\zeta_1=x_2+ix_1\quad\text{and}\quad \zeta_2=x_4+ix_3\, .
\end{equation}
Then the function $H$ becomes
\begin{equation}
H(\zeta_1,\zeta_2)=-\frac{\omega_1}{2}\zeta_1\bar{\zeta}_1-\frac{\omega_2}{2}\zeta_2\bar{\zeta}_2\, ,
\end{equation}
and using 
\begin{eqnarray}
\frac{\partial}{\partial x_1}&=&\frac{\partial \zeta_1}{\partial x_1}\frac{\partial}{\partial \zeta_1}+\frac{\partial \bar{\zeta}_1}{\partial x_1}\frac{\partial}{\partial \bar{\zeta}_1}=i\frac{\partial}{\partial \zeta_1}-i\frac{\partial}{\partial \bar{\zeta}_1}\\
\frac{\partial}{\partial x_2}&=&\frac{\partial \zeta_1}{\partial x_2}\frac{\partial}{\partial \zeta_1}+\frac{\partial \bar{\zeta}_1}{\partial x_2}\frac{\partial}{\partial \bar{\zeta}_1}=\frac{\partial}{\partial \zeta_1}+\frac{\partial}{\partial \bar{\zeta}_1}
\end{eqnarray}
for $(x_1,x_2)$, the operator $\{H\}$ becomes 
\begin{equation}
\partial_{x_2}H\partial_{x_1}-\partial_{x_1}H\partial_{x_2}=2i\left(\partial_{\zeta_1}H\partial_{\bar{\zeta}_1}-\partial_{\bar{\zeta}_1}H\partial_{\zeta_1}\right)
=i\omega_1\left(\bar{\zeta}_1\partial_{\bar{\zeta}_1}-\zeta_1\partial_{\zeta_1}\right)\, ,
\end{equation}
with a similar expression holding for $(x_3,x_4)$.  Hence for any
$\vec{n}=(n_1,n_2)\in\mathbb{N}^2$ and $\vec{m}=(m_1,m_2)\in\mathbb{N}^2$ we
obtain
\begin{equation}
\label{opHcomplex}
  \{H\}{\zeta}^{\vec{n}}
  \bar{{\zeta}}^{\vec{m}}=i\left(\omega_1m_1-\omega_1n_1+\omega_2m_2-\omega_2n_2\right){\zeta}^{\vec{n}}
  \bar{{\zeta}}^{\vec{m}}=i\vec{\omega}\cdot(\vec{m}-\vec{n}){\zeta}^{\vec{n}} 
  \bar{{\zeta}}^{\vec{m}}\, ,
\end{equation}
where we introduced the vector $\vec{\omega}=(\omega_1,\omega_2)$ and we used
the compact notation ${\zeta}^{\vec{n}}=\zeta_1^{n_1}\zeta_2^{n_2}$, for the
complex vector ${\zeta}=(\zeta_1,\zeta_2)$.

We note that from the knowledge of the operators' action on such monomials
$\zeta^{\vec{n}}\bar{\zeta}^{\vec{m}}$, we can reconstruct the operator action
on any regular function. The operators are linear and they will be
applied on polynomials in the $\vec{x}$ variable, which are nothing more than
polynomials in the complex variables.

Let us now compute the time-$1$ flow of $\{H\}$ by using complex variables.
Starting from ~\eqref{opHcomplex} and then proceeding by induction, we can
easily prove that for all $k\in\mathbb{N}$
\begin{equation}
\{H\}^k\zeta^{\vec{n}}\bar{\zeta}^{\vec{m}}=\left(i\vec{\omega}\cdot(\vec{m}-\vec{n})\right)^k\zeta^{\vec{n}}\bar{\zeta}^{\vec{m}}\, ,
\end{equation}
and finally
\begin{equation}
e^{\{H\}}{\zeta}^{\vec{n}}\bar{\zeta}^{\vec{m}}=\sum_{k=0}^{\infty}\frac{\left(i\vec{\omega}\cdot(\vec{m}-\vec{n})\right)^k}{k!}{\zeta}^{\vec{n}}\bar{{\zeta}}^{\vec{m}}=e^{i\vec{\omega}\cdot(\vec{m}-\vec{n})}{\zeta}^{\vec{n}}\bar{\zeta}^{\vec{m}}.
\end{equation}
Similarly
$e^{-\{H\}}{\zeta}^{\vec{n}}\bar{{\zeta}}^{\vec{m}}=e^{-i\vec{\omega}\cdot(\vec{m}-\vec{n})}{\zeta}^{\vec{n}}\bar{{\zeta}}^{\vec{
m}}$.

Assuming a non-resonance condition
\begin{equation}
  \vec{\omega}\cdot(\vec{m}-\vec{n})\neq 2k\pi\quad \forall
  \vec{n}\neq\vec{m}\in\mathbb{N}^2\setminus \{0\}\quad \text{and}\quad
  \forall k\in \mathbb{Z}\, , 
\end{equation}
a possible choice for the operator $\mathcal{G}$ is the following
\begin{equation}
\label{G}
\mathcal{G}{\zeta}^{\vec{n}}\bar{{\zeta}}^{\vec{m}}:=\frac{1}{1-e^{-i\vec{\omega}\cdot(\vec{m}-\vec{n})}}\mathcal{N}{\zeta}^{\vec{n}}\bar{{\zeta}}^{\vec{m}}
\end{equation}
with 
\begin{equation}
  \mc{N}{\zeta}^{\vec{n}}\bar{{\zeta}}^{\vec{m}}=\begin{cases}{\zeta}^{\vec{n}}\bar{{\zeta}}^{\vec{m}}
    & \text{if $\vec{\omega}\cdot(\vec{m}-\vec{n})\neq 2\pi k$}\\
0 & \text{otherwise} 
\end{cases}\, .
\end{equation}
It is easy to check that the operator $\mc{G}$ defined by (\ref{G}) verifies
(\ref{propG}). In order to do so we introduce the compact notation
\begin{equation}
Z_{n,m}:={\zeta}^{\vec{n}}\bar{{\zeta}}^{\vec{m}}\quad \text{and}\quad
\square_{n,m}:=e^{-i\vec{\omega}\cdot(\vec{m}-\vec{n})}\, .
\end{equation}
Developing the left hand side of~\eqref{propG} and using the linearity of all
operators, we get
\begin{eqnarray}
\mc{G}\left(1-e^{-\{H\}}\right)\mc{G}Z_{n,m}&=&
\left(\mc{G}-\mc{G}e^{-\{H\}}\right)\frac{1}{1-\square_{n,m}}Z_{n,m}\\
&=&\frac{1}{1-\square_{n,m}}\mc{G}Z_{n,m}-\frac{1}{1-\square_{n,m}}\mc{G}e^{-\{H\}}Z_{n,m}\\
&=&\frac{1}{(1-\square_{n,m})^2}Z_{n,m}-\frac{1}{1-\square_{n,m}}\mc{G}(\square_{n,m} Z_{n,m})\\
&=&\frac{1}{(1-\square_{n,m})^2}Z_{n,m}-\frac{\square_{n,m}}{(1-\square_{n,m})^2}Z_{n,m}\\
&=&\frac{1}{1-\square_{n,m}}Z_{n,m}=\mc{G}Z_{n,m}\, .
\end{eqnarray}

\subsection{Expression of the control term $F_2$}
The function $F_2$ is defined by~\eqref{eq:f2}, where $V$ is a known
function. The term $\mc{G}V$ will be computed starting from the previously
obtained expression of $\mc{G}$. To construct $F_2$ we first have to express
$V$ in the complex variables (\ref{complexCoor}):
\begin{eqnarray}
 V(\zeta_1,\zeta_2)=-\frac{1}{24}i\zeta_1^3
+\frac{1}{8}i\bar{\zeta}_1\zeta_1^2
-\frac{1}{8}i\bar{\zeta}_1^2\zeta_1
+\frac{1}{24}i\bar{\zeta}_1^3
+\frac{1}{8}i\zeta_1\zeta_2^2
-\frac{1}{4}i\zeta_1\zeta_2\bar{\zeta}_2\nonumber\\
+\frac{1}{8}i\zeta_1\bar{\zeta}_2^2
-\frac{1}{8}i\bar{\zeta}_1\zeta_2^2
+\frac{1}{4}i\bar{\zeta}_1\bar{\zeta}_2\zeta_2
-\frac{1}{8}i\bar{\zeta}_1\bar{\zeta}_2^2\, .
\label{eq:53}
\end{eqnarray}

By the linearity of the operator, and by using (\ref{G}), we easily compute
$\mc{G}V$. In particular we apply $\mc{G}$ to each term of (\ref{eq:53}). Then
using the inverse change of coordinates
\begin{equation}
x_1=\frac{1}{2}i(\bar{\zeta}_1-\zeta_1)\quad\text{and}\quad
x_2=\frac{1}{2}(\zeta_1+\bar{\zeta}_1)\, 
\end{equation}
(similar expressions hold for $(x_3,x_4)$ and $(\zeta_2,\bar{\zeta}_2)$), we
can go back to the original variables $\vec{x}$. The obtained expression after
some algebraic simplifications is
\begin{eqnarray}
\mc{G}V&=&-1/6\,\csc \left( 3/2\,\omega_1 \right) \left[ x_2\,\cos \left(
1/2\,\omega_1 \right) +x_1\,\sin \left( 1/2\,\omega_1 \right) \right] \left[
{x_1}^{2}-3\,{x_3}^{2}+{x_2}^{2}\right.\nonumber\\
&&\left.+ \left(
2\,{x_1}^{2}-6\,{x_3}^{2}+{x_2}^{2} \right) \cos \left( \omega_1 \right)
-x_1\,x_2\,\sin \left( \omega_1 \right) \right] \nonumber\\
&&+1/4\,{\frac {\sin \left(
\omega_2 \right) }{\cos \left( \omega_1-\omega_2 \right) -\cos \left( \omega_2
\right) }}\nonumber\\
&&-1/4\,{\frac { \left(
-x_2\,{x_4}^{2}+x_2\,{x_3}^{2}+2\,x_1\,x_4\,x_3 \right) \sin \left( \omega_2
\right) }{\cos \left( \omega_2 \right) -\cos \left( \omega_1+\omega_2 \right)
}}
\end{eqnarray}

Then the explicit expression of the control term $F_2$ is
\begin{eqnarray}
F_2&=&1/2 ( x_1^2-x_3^2 ) \Bigg( -1/6\csc ( 3/2\omega_1 ) \cos ( 1/2\omega_1 )
( x_1^2-3x_3^2+x_2^2\nonumber\\
&& +( 2x_1^2-6x_3^2+x_2^2 ) \cos ( \omega_1 )
-x_1x_2\sin ( \omega_1 ) ) \nonumber\\
&&-1/6\csc ( 3/2\omega_1 ) \Big( x_2\cos (
1/2\omega_1 ) +x_1\sin ( 1/2\omega_1 ) \Big) ( 2x_2+2x_2\cos ( \omega_1 )
-x_1\sin ( \omega_1 ) )\nonumber \\
&&-1/4{\frac { ( x_3^2-x_4^2 ) \sin ( \omega_2 )
}{\cos ( \omega_2 ) -\cos ( \omega_1+\omega_2 ) }} \Bigg) +1/4{\frac {x_1x_3 (
2x_1x_3-2x_2x_4 ) \sin ( \omega_2 ) }{\cos ( \omega_2 ) -\cos (
\omega_1+\omega_2 ) }}.
\end{eqnarray}

\section*{Acknowledgements}

Numerical simulations were made on the local computing resources (Cluster
URBM-SYSDYN) at the University of Namur (FUNDP, Belgium).


\end{document}